# Revisiting thermodynamics in (LiF, NaF, KF, $CrF_2$)-$CrF_3$ by first-principles calculations and CALPHAD modeling

Rushi Gong[1], Shun-Li Shang[1], Yi Wang[1], Jorge Paz Soldan Palma[1], Hojong Kim[1], and Zi-Kui Liu[1]

[1]Department of Materials Science and Engineering, The Pennsylvania State University, University Park, PA 16802, USA

**Abstract**

The thermodynamic description of the (LiF, NaF, KF, $CrF_2$)-$CrF_3$ systems has been revisited, aiming for a better understanding of the effects of Cr on the FLiNaK molten salt. First-principles calculations based on density functional theory (DFT) were performed to determine the electronic and structural properties of each compound, including the formation enthalpy, volume, and bulk modulus. DFT-based phonon calculations were carried out to determine the thermodynamic properties of compounds, for example, enthalpy, entropy, and heat capacity as functions of temperature. Phonon-based thermodynamic properties show a good agreement with experimental data of binary compounds LiF, NaF, KF, $CrF_3$, and $CrF_2$, establishing a solid foundation to determine thermodynamic properties of ternary compounds as well as to verify results estimated by the Neumann-Kopp rule. Additionally, DFT-based ab initio molecular dynamics (AIMD) simulations were employed to predict mixing enthalpies of liquid salts. Using DFT-based results and experimental data in the literature, the (LiF, NaF, KF, $CrF_2$)-$CrF_3$ system has been remodeled in terms of the CALculation of PHAse Diagrams (CALPHAD) approach using the modified quasichemical model with quadruplet approximation (MQMQA) for liquid. Calculated phase stability in the present work shows an excellent agreement with experiments, indicating the effectiveness of combining DFT-based total energy, phonon, and AIMD calculations, and CALPHAD modeling to provide the thermodynamic description in complex molten salt systems.

## Highlights

- Temperature-dependent thermodynamic properties of compounds in (LiF, NaF, KF, $CrF_2$)-$CrF_3$ predicted by DFT-based phonon calculations.
- Mixing enthalpies of liquid salts in (LiF, NaF, KF)-$CrF_3$ predicted by AIMD.
- The (LiF, NaF, KF, $CrF_2$)-$CrF_3$ system re-modeled with MQMQA to describe liquid.

## 1 Introduction

Molten Salt Reactor (MSR) is one of the few game-changing concepts with rigorous safety standards while simultaneously achieving high levels of reliability and efficiency due to the use of liquid, instead of solid, fuel in the reactor [1–3]. The MSR concept consists of a molten salt mixture (such as $LiF$-$BeF_2$-$UF_4$), in which the fissile and fertile isotopes (such as $^{233}U$, $^{235}U$, $^{238}U$, and/or $^{239}Pu$) are dissolved, circulating from the reactor core to the heat exchanger continuously [1,4]. A very important feature of this system is its safety, requiring more attention to select molten salts [5]. Alkali and alkaline-earth metal fluorides, in which actinide fluorides (such as $UF_4$ and $PuF_3$) can be dissolved, are considered as the key materials for MSR fuel salts [1]. For example, the fuel salt of $^7LiF$-$BeF_2$-$ZrF_4$-$UF_4$ with $^{235}U$, $^{233}U$, and/or $^{239}Pu$ as the fissile driver was used in the Molten Salt Reactor Experiment (MSRE) operated at Oak Ridge National Laboratory (ORNL) from 1965 to 1969 [1] with the coolant salt in the secondary loop being $^7Li_2BeF_4$. The Molten Salt Chemistry Workshop [1] suggested that fluoride salts, which are similar to the MSRE salts, are primarily for the thermal spectrum applications and for the primary and secondary coolant candidates, such as the FLiNaK, i.e., the LiF-NaF-KF eutectics with its mole fraction around 0.465-0.115-0.420 [6,7].

Chromium (Cr) is one of the key elements in current reference structural materials for fluoride salt reactors, for example, the Hastelloy-N as the container material (72%Ni-16%Mo-7%Cr-5%Fe in wt.%) [8,9]. In the fluoride salt environment, Cr in Ni-based alloy is more susceptible to corrosion compared to other metal elements [10–12]. For example, Olson et al. [10] performed corrosion tests on a number of Ni-based alloys with different Cr alloying contents. They showed the dissolution of Cr into molten salt and correlated the relationship between Cr content and corrosion resistance [10]. Ouyang et al. [11] investigated the corrosion behavior of Hastelloy-N in FLiNaK after 100-1000 h at 700 °C, and the aggregate dissolution of Cr was observed. Recently, Liu et al. [12] studied Hastelloy-N and showed that the corrosion rate of Cr in FLiNaK-$CrF_3$ is higher than that in FLiBe-$CrF_3$ with $Cr^{3+}$ as the product. Thus, it is important to investigate solubility and multivariate distribution patterns of Cr in FLiNaK. There exists extensive thermodynamic modeling work related to the LiF-NaF-KF system, several studies, including those by Chartrand and Pelton [13], Wang et al. [14], and Ard et al. (MSTDB-TC database) [15] have explored this system using thermodynamic modeling. However, we noticed that thermodynamic modeling of the FLiNaK-$CrF_3$ system has been performed based on incomplete and empirically estimated thermochemical data by Yin et al. [16,17] and Dumaire et al. [18] (see Sec. 2.1). A comprehensive study on thermodynamic properties of all solid and liquid phases in FLiNaK-$CrF_3$ is yet to be conducted.

In the present work, the thermodynamic description of the (LiF, NaF, KF, $CrF_2$)-$CrF_3$ system is re-modeled using the CALPHAD (CALculation of PHAse Diagram) approach [19–21] with experimental data in the literature, and thermochemical data from the present density functional theory (DFT) based first-principles calculations, phonon calculations, and ab initio molecular dynamics (AIMD) simulations. In particular, the modified quasichemical model with quadruplet approximation (MQMQA) [22] is adopted for the liquid phase. The open-source software ESPEI (Extensible Self-optimizing Phase Equilibria Infrastructure) [23,24] with the computational engine of PyCalphad [25,26] is used for the present CALPHAD modeling. A comprehensive description of thermodynamic properties in the (LiF, NaF, KF, $CrF_2$)-$CrF_3$ system is achieved and compared well with experimental data in the literature.

## 2 Literature review of thermodynamic properties

### 2.1 Thermochemical data

The (LiF, NaF, KF, $CrF_2$)-$CrF_3$ system includes five binary (endmember) compounds, i.e., LiF, NaF, KF, $CrF_3$ and $CrF_2$, and eight ternary (intermetallic) compounds, i.e., $Li_3CrF_6$, $Na_3CrF_6$, $Na_5Cr_3F_{14}$, $NaCrF_4$, $K_3CrF_6$, $K_2CrF_5$, $KCrF_4$, and $K_2Cr_5F_{17}$. These ternary compounds were first suggested by De Kozak [27] and confirmed by structural studies [28–35] via the X-ray diffraction (XRD) method, which was summarized by Dumaire et al. [18]. However, thermochemical data of these compounds are scarce. Yin et al. [16,17,36] performed DFT calculations at 0 K to determine the formation enthalpies of $Li_3CrF_6$, $Na_3CrF_6$, $Na_5Cr_3F_{14}$, $NaCrF_4$, and $KCrF_4$. Dumaire et al. [18] estimated the heat capacities of these compounds based on the Neumann-Kopp rule in terms of the compositional average of heat capacity values of the corresponding compounds or elements [37]. For the liquid phase, experimental data such as mixing enthalpy are not available in the AF-$CrF_3$ (A=Li, Na, and K) systems. Instead, Yin et al. [16,17,36] applied an empirical model to estimate the mixing enthalpy of liquid from the parameters of ions such as ionic radius.

## 2.2 Phase equilibrium data

Phase equilibria in the LiF-$CrF_3$, NaF-$CrF_3$, and KF-$CrF_3$ binary systems were investigated by De Kozak [27,28] using differential thermal analysis (DTA). In LiF-$CrF_3$, two eutectic reactions were measured, i.e., Liquid ↔ LiF + $Li_3CrF_6$ at 1003 K and around mole fraction X($CrF_3$) = 0.15 and Liquid ↔ $CrF_3$ + $Li_3CrF_6$ at 1059 K and X($CrF_3$) = 0.35. In NaF-$CrF_3$, one peritectic reaction of Liquid + $CrF_3$ ↔ $NaCrF_4$ at 1234 K and three eutectic reactions were determined, i.e., Liquid ↔ $NaCrF_4$ + $Na_5Cr_3F_{14}$ at 1133 K, Liquid ↔ $Na_3CrF_6$ + $Na_5Cr_3F_{14}$ at 1143 K, and Liquid ↔ $Na_3CrF_6$ + NaF at 1166 K and around X($CrF_3$) = 0.123. In KF-$CrF_3$, De Kozak [27,28] reported three peritectic reactions and two eutectic reactions, i.e., Liquid + $CrF_3$ ↔ $K_2Cr_5F_{17}$ at 1390 K, Liquid + $K_3CrF_6$ ↔ $K_2CrF_5$ at 1133 K, and Liquid + $K_2Cr_5F_{17}$ ↔ $KCrF_4$ at 1200 K, and Liquid ↔ $K_3CrF_6$ + KF at 1115 K and around X($CrF_3$) = 0.048, and Liquid ↔ $K_2CrF_5$ + $KCrF_4$ at 1112 K and around X($CrF_3$) = 0.45. Sturm [30] reported phase equilibria in $CrF_2$-$CrF_3$ via quenching experiments and suggested one solid solution phase in $CrF_2$-$CrF_3$ with composition of $CrF_3$ between 0.42 and 0.46 (near $Cr_2F_5$). However, the stability of this $Cr_2F_5$ solid solution phase was not explored in temperatures below 1023 K. The melting point of $Cr_2F_5$ was determined to be around 1270 K [30]. Sturm [30] reported one eutectic reaction, Liquid ↔ $CrF_2$ + $Cr_2F_5$ at 1103 K around X($CrF_3$) = 0.14, and one peritectic reaction Liquid + $CrF_3$ ↔ $Cr_2F_5$ at 1272 K around X($CrF_3$) = 0.29. Two solid solution phases near the endmembers $CrF_3$ and $CrF_2$ were identified from X($CrF_3$) = 0~0.01 and from X($CrF_3$) = 0.90 ~1, respectively.

# 3 Methodology

## 3.1 DFT-based first-principles calculations

### 3.1.1 *Helmholtz energy at finite temperatures*

The Helmholtz energy $F(V,T)$ as a function of volume ($V$) and temperature ($T$) in terms of the DFT-based quasiharmonic approximation (QHA) can be determined by [38],

$$F(V,T) = E(V) + F_{el}(V,T) + F_{vib}(V,T) \qquad \textbf{\textit{Eq. 1}}$$

where the first term $E(V)$ is static energy at 0 K without the zero-point vibrational energy. In the present work, a four-parameter Birch-Murnaghan (BM4) equation of state (EOS) [38] as shown in ***Eq. 2*** was used to obtain equilibrium properties at zero external pressure ($P$ = 0 GPa), including the static energy $E_0$, volume ($V_0$) , bulk modulus ($B_0$) and its pressure derivate ($B$').

$$E(V) = a + bV^{-2/3} + cV^{-4/3} + dV^{-2} \qquad \textbf{\textit{Eq. 2}}$$

where $a$, $b$, $c$, and $d$ are fitting parameters. The second term in ***Eq. 1*** , $F_{el}(V,T)$, represents the temperature-dependent thermal electronic contribution [39],

$$F_{el}(V,T) = E_{el}(V,T) - T \cdot S_{el}(V,T) \qquad \textbf{\textit{Eq. 3}}$$

where $E_{el}$ and $S_{el}$ are the internal energy and entropy of thermal electron excitations, respectively, which can be obtained by electronic density of states (DOS). Note that the thermal electronic contribution to Helmholtz free energy is negligible for non-metal, considering the Fermi level lies in the band gap. The third term in ***Eq. 4***, $F_{vib}(V,T)$, represents the vibrational contribution [39,40],

$$F_{vib}(V,T) = k_B T \sum_q \sum_j \ln\left\{2\sinh\left[\frac{\hbar\omega_j(q,V)}{2k_B T}\right]\right\} \quad \textbf{Eq. 5}$$

where $\omega_j(q,V)$ represents the frequency of the $j^{\text{th}}$ phonon mode at wave vector $q$ and volume $V$, and $\hbar$ the reduced Plank constant.

### *3.1.2 Details of first-principles calculations*

All DFT-based first-principles and phonon calculations in the present work were performed by the Vienna *Ab initio* Simulation Package (VASP) [41]. The projector augmented-wave method (PAW) was used to account for electron-ion interactions in order to increase computational efficiency compared with the full potential methods [42,43]. Electron exchange and correlation effects were described using both the local density approximation (LDA) [44] and the generalized gradient approximation (GGA) as implemented by Perdew, Burke, and Ernzerhof (PBE) [45]. In addition, the DFT+U approach was employed for 11 compounds containing Cr, i.e., $CrF_2$, $CrF_3$, $Cr_2F_5$, $Li_3CrF_6$, $Na_3CrF_6$, $Na_5Cr_3F_{14}$, $NaCrF_4$, $K_3CrF_6$, $K_2CrF_5$, $KCrF_4$, $K_2Cr_5F_{17}$. The effective U values for Cr were selected as 3.7 eV, considering 3 ~ 4eV was commonly used in the literature [46–48]. The spin configurations were also considered for these 11 compounds containing Cr. All possible configurations by varying spin up and spin down of Cr atoms were explored by the ATAT code [49]. The spin configuration with the lowest energy for each Cr-containing compound was used for DFT and phonon calculations.

In the present work, DFT-based first-principles and phonon calculations were performed by using the open-source software DFTTK [50]. Using DFTTK, structure information is the only required input, then robust relaxation schemes can be automatically performed to obtain equilibrium results at 0 K and thermodynamic properties at finite temperatures through the QHA. During DFTTK calculations, the plane-wave cutoff energy was set as 520 eV. Table 1 lists the k-points meshes for DFT-based total energy calculations, the supercell sizes, and the k-points meshes for phonon calculations. The phonon DOS and force constants were analyzed using the YPHON code [51], which has been integrated into DFTTK [50].

## 3.2 AIMD simulations

The *ab initio* molecular dynamics (AIMD) simulations in the present work were also performed by VASP [41]. The supercells containing 108 or 128 atoms with periodic boundary were used for at least six different compositions in the AF-$CrF_3$ (A= Li, Na, and K) systems, including $A_{64}F_{64}$, $A_{42}Cr_6F_{60}$, $A_{36}Cr_9F_{63}$, $A_{32}Cr_{16}F_{80}$, $A_{18}Cr_{18}F_{72}$, $A_{16}Cr_{24}F_{88}$, $A_{10}Cr_{22}F_{76}$, and $Cr_{32}F_{96}$ (A=Li, Na, and K). The NVT canonical ensemble (i.e., the fixed number of atoms N, volume V, and temperature T) with a Nosé thermostat for temperature control [52] were employed in the present work. The temperature for each supercell was set as 1700 K, which is above all the temperatures of liquidus in the AF-$CrF_3$ (A= Li, Na, and K) systems. A single Γ point 1×1×1 was used as the k-point mesh, together with a 400 eV cutoff energy. During AIMD simulations, the Newton's equation of motion was solved via the Verlet algorithm with a time step of 2 fs and each calculation is run for 10,000 steps to reach thermal equilibrium.

## 3.3 CALPHAD modeling

In the present work, the compounds and endmembers in the AF-$CrF_3$ (A=Li, Na, and K) systems are treated as stoichiometric compounds, including four binary endmembers, i.e., LiF, NaF, KF, and $CrF_3$, and eight ternary compounds, i.e., $Li_3CrF_6$, $Na_3CrF_6$, $Na_5Cr_3F_{14}$, $NaCrF_4$, $K_3CrF_6$, $K_2CrF_5$, $KCrF_4$, and $K_2Cr_5F_{17}$ (as listed in Sec.2.1). Thermodynamic functions of the binary endmembers are taken from the JRC database [53], JANAF tables [54], IVTAN tables [55], and SSUB database [56]. The Gibbs energy can be expressed as

$$G_m = \Delta_f H_m^0(298.15) - T\,S_m^0(298.15) + \int_{298.15}^{T} C_{P,m} dT - T\int_{298.15}^{T} \frac{C_{P,m}}{T} dT \qquad \textbf{\textit{Eq. 6}}$$

where $\Delta_f H_m^0(298.15)$ is the standard formation enthalpy, $S_m^0(298.15)$ the standard entropy at 298.15 K, and $C_{P,m}$ the heat capacity. For ternary compounds, their thermodynamic data including enthalpy, entropy, and heat capacity are obtained through DFT-based first-principles and phonon calculations, which are presented in Sec. 4.1.

The MQMQA [57,58] is used to describe the liquid phase by considering the short-range ordering (SRO) occurred in liquid salts. Within the MQMQA, two sublattices which separate the cations from the anions are introduced to form quadruplets and account for the first nearest neighbors (FNN) and the second nearest neighbors (SNN) interactions. Here, the model (A, Cr)(F) is introduced and hence the AA/FF, CrCr/FF, and ACr/FF quadruplets (A=Li, Na, and K) are formed to consider the interactions among them. Coordination numbers Z are defined to describe the quadruplets. Z of anions can be calculated from ***Eq. 7*** to maintain charge neutrality as follows

$$\frac{q_{\mathrm{A}}}{Z_{\mathrm{AB:FF}}^{\mathrm{A}}}+\frac{q_{\mathrm{B}}}{Z_{\mathrm{AB:FF}}^{\mathrm{B}}}=2\times\frac{q_{\mathrm{F}}}{Z_{\mathrm{AB:FF}}^{\mathrm{F}}} \qquad \textbf{\textit{Eq. 7}}$$

where $q_i$ is the charges of ion $i$ (= Li, Na, K, Cr, or F). Coordination numbers used in the present work are shown in Table 2**.**

The excess Gibbs energy $G^{excess}$ relates to the formation Gibbs energy of the quadruplet, $\Delta g_{quadruplet}^{ex}$, by considering the following reaction,

$$(\mathrm{A}-\mathrm{F}-\mathrm{A})+(\mathrm{Cr}-\mathrm{F}-\mathrm{Cr})=2(\mathrm{A}-\mathrm{F}-\mathrm{Cr}) \qquad \Delta g_{\mathrm{Acr:F_2}}^{ex} \qquad \textbf{\textit{Eq. 8}}$$

where $\Delta g_{\mathrm{Acr:F_2}}^{ex}$ represents the Gibbs energy change when forming the quadruplet and can be described by

$$\Delta g_{\mathrm{Acr:F_2}}^{ex}=\sum_{i\geq 0,j\geq 0} g_{\mathrm{Acr:F}}^{ij}\chi_{\mathrm{Acr:F}}^{i}\chi_{\mathrm{CrA:F}}^{j} \qquad \textbf{\textit{Eq. 9}}$$

where $g_{\mathrm{Acr:F_2}}^{ij}$is a function of temperature and it is independent of composition. $\chi_{\mathrm{Acr:F_2}}^{i}$ and $\chi_{\mathrm{CrA:F_2}}^{j}$ are composition-dependent terms,

$$\chi_{\mathrm{Acr:F}} = \frac{n_{\mathrm{AA:F}}}{n_{\mathrm{Acr:F}} + n_{\mathrm{AA:F}} + n_{\mathrm{CrCr:F}}} \qquad \textbf{\textit{Eq. 10}}$$

where $n_{\mathrm{Acr:F}}$ is the moles of $(\mathrm{A}-\mathrm{F}-\mathrm{Cr})$ shown in ***Eq. 10***.

In the $CrF_2$-$CrF_3$ system, there are three solid solution phases, i.e., $CrF_2$ near the Cr-rich region, $CrF_3$ near the F-rich region, and $Cr_2F_5$ showing on the middle region of the $CrF_2$ – $CrF_3$ phase diagram. The present work adopts the same models used by Dumaire et al. [18], where the regular solution model with the Kohler-Toop interpolation [59–62] is used for $Cr_2F_5$. For solid solution phases near $CrF_2$ and $CrF_3$, the sublattice model is used for each phase, respectively, considering the Wyckoff positions of $CrF_2$ and $CrF_3$ as follows. $CrF_2$ possesses the symmetry with space group $P2_1/c$ with two Wyckoff sites of 2a and 4e, and the sublattice model $(Cr, Va)_1(F, Va)_2$ is hence used with Va representing the vacancy. $CrF_3$ phase is modeled by $(Cr, Va)_1(F, Va)_3$ by considering its space group $R\bar{3}c$ and the two Wyckoff sites of 2b and 6e. The Gibbs energy per mole of xxx is formulated as:

$$G_m = \sum_{i=Cr,Va} \sum_{j=F,Va} y_i' y_j'' \, {}^{o}G_{i:j} + RT\left( \sum_{i=Cr,Va} y_i' \ln(y_i') + \sum_{j=F,Va} y_j'' \ln(y_j'') \right) + y_{Cr}' y_{Va}' \left( y_F'' L_{Cr,Va:F} \right) + y_F'' y_{Va}'' \left( y_{Cr}' L_{Cr:F,Va} \right) \qquad \textbf{\textit{Eq. 11}}$$

where $y_i^{(s)}$ is the site fraction of component $i$ on sublattice $s$, ${}^{o}G_{i:j}$ the Gibbs energy of the endmember $(i{:}j)$, and $L$ the interaction parameters which can be expanded using the Redlich-Kister polynomials [63].

Machine learning (ML) was used to estimate more phase equilibria data in terms of a graphic neural network model developed by Hong et al. [64] to predict melting points of compounds with composition as input. Melting temperatures of the present ternary compounds including $Li_3CrF_6$, $Na_3CrF_6$, $Na_5Cr_3F_{14}$, $NaCrF_4$, $K_3CrF_6$, $K_2CrF_5$, $KCrF_4$, and $K_2Cr_5F_{17}$ are estimated by this ML model [64].

Thermodynamic modeling of the (LiF, NaF, KF, $CrF_2$)-$CrF_3$ system was carried out by means of the open-source software ESPEI [23], which uses PyCalphad [25] as computational engine for thermodynamic calculations with the newly implemented MQMQA [65]. All model parameters were simultaneously optimized through the Bayesian approach using the Markov Chain Monte Carlo (MCMC) method [23]. The input data were primarily experimental phase equilibrium data including two or more co-existing phases. For stochiometric compounds, their thermochemical data from DFT-based calculations were also used as input. For the liquid phase, its mixing enthalpy from AIMD calculations was used as input for refining model parameters. In the present work, each model parameter employed two Markov chains with a standard derivation of 0.1 when initializing its Gaussian distribution. During the modeling process, the chain values can be tracked and the MCMC processes were performed until the model parameters converged.

# 4 Results and discussion

## 4.1 Thermodynamic properties in (LiF, NaF, KF, $CrF_2$)-$CrF_3$ by first-principles calculations

Table 3 shows the predicted equilibrium properties of $V_0$, $B_0$, and B' by the EOS E-V fitting at 0 K in comparison with experimental bulk moduli [66–71]. The $B_0$ results from GGA show a good agreement with experimental measurements. For the LiF compound, GGA predicts $B_0$ = 67.6 GPa and B' = 4.17, aligning well with the measured values of 65.4 GPa and 4.98 by Boehler et al. [68]. In comparison with the three measured $B_0$ values of 66.5 GPa by Yagi [66], 76.9 GPa by Haussühl [67], and 65.4 GPa by Boehler et al. [68], the $B_0$ result by GGA shows a mean absolute error (MAE) of 4.2 GPa, while LDA predicts $B_0$ = 86.5 GPa with a higher MAE of 16.9 GPa. Considering the NaF compound, GGA predicts $B_0$ = 45.0 GPa, matching with the measured 45.9 GPa by Yagi [66] but lower than the measured values of 53.8 GPa by Haussühl [67], 52.3 GPa Rao [69], and 48.2 GPa Bensch et al. [70] with the MAE around 5 GPa. On the other hand, LDA predicts $B_0$ = 61.4 GPa, which is higher than the experimental $B_0$ [66,67,69,70] with MAE = 11.4 GPa. Additionally, for the B' values of NaF, LDA predicts B' = 4.74, which is higher than the predicted 4.60 by GGA and closer to 5.89 reported by Bensch et al. [70]. Regarding the KF compound, the measured $B_0$ values (37.0 GPa by Yagi [66] and 35.5 GPa by Haussühl [67]) fall between the LDA result of 43.5 GPa and the GGA result of 28.9 GPa. The MAE value of LDA with respect to the measured B0 values is 7.25 GPa, slightly lower than MAE = 7.35 GPa by GGA. As for the $CrF_3$ compound, GGA+U reports $B_0$ = 29.3 GPa, showing a 0.3% difference compared to 29.2 GPa measured by Jørgensen et al. [71]. LDA+U predicts $B_0$ = 46.2 GPa, which is 58% higher than the measured 29.2 GPa [71]. The present results indicate that LDA predicts smaller $V_0$ and higher $B_0$ values than those from GGA. It is consistent with the previous observations that LDA tends to underestimate lattice constants and overestimate cohesive energy with respect to GGA [72,73].

Figure 1 compares the phonon DOS of LiF, NaF, and KF obtained using LDA and GGA in comparison with direct measurements by neutron scattering [74,75] or fittings in terms of measurements [76]. Overall, the peak positions of the experimental phonon DOS are well reproduced by both LDA and GGA. However, in the low-frequency region (e.g., < 5THz for LiF and NaF, and < 3THz for KF), the phonon DOS predicted by LDA show a better match in both the shape and the peak position with respect to experimental data [74–76] than those from GGA. This observation suggests that LDA predicts more reliable thermodynamic properties of LiF, NaF, and KF when employing the phonon based QHA, since these properties are mainly regulated by phonon DOS at low frequency regions [77].

Figure 2 illustrates a comparison of the predicted heat capacity ($C_p$), entropy (S), and enthalpy (H-$H_{300}$) of LiF, NaF, and KF in terms of the phonon-based QHA, where the DFT calculations were conducted using both the LDA and GGA, and the predicted results are compared to the data from the SSUB database [56]. In general, thermodynamic properties predicted by LDA align well with the results from SSUB. The most substantial difference between LDA and SSUB is the $C_p$ values of KF, where a 6% disparity is noted. Furthermore, these comparisons reveal that the LDA results exhibit a better agreement with SSUB than the GGA results, particularly for LiF. For example at 1100 K, the $C_p$ values predicted by LDA demonstrate only a 2% difference compared to those from the SSUB [56], whereas an 18% difference is observed when using GGA. Figure 2 indicates that the QHA in terms of LDA yields more reliable predictions of thermodynamic properties in LiF-NaF-KF-based system, agreeing with the observations in phonon DOS in Figure 1. Subsequent DFT calculations were hence performed using the LDA approach.

Figure 3 shows the present DFT predictions and the values in SSUB [56] regarding $C_p$, S, and H-$H_{300}$ for $CrF_3$ and $CrF_2$. In general, the present predictions tend to be lower than those obtained from SSUB [56]. For $CrF_3$, the DFT predicted $C_p$ = 19.77 J/mol-atom-K closely aligns with the SSUB value of 19.73 J/mol-atom-K at 300 K. As the temperature increases to 1300 K, the difference increases to 9%. Similarly, for $CrF_2$, the $C_p$ = 21.22 J/mol-atom-K at 300 K by DFT is in good agreement with the value of 21.66 from SSUB. At a higher temperature of 1140 K, the difference expands to 12%. Regarding entropy, DFT predicts lower S values for both $CrF_3$ and $CrF_2$ than those in SSUB across the temperature range shown in Figure 3. At high temperatures (e.g., 1600 K for $CrF_3$ and 1140 K for $CrF_2$), it shows a 10% difference in $CrF_3$ and 14% in $CrF_2$. It is found that a good agreement is observed regarding H-$H_{300}$ values of $CrF_3$ and $CrF_2$ between the DFT calculations and the SSUB at lower temperatures (< 1000 K for $CrF_3$ and < 600 K for $CrF_2$). H-$H_{300}$ from DFT becomes slightly lower than that in SSUB [56] with increasing temperature, with the differences, for example, around 5% in $CrF_3$ at 1650 K and 9% in $CrF_2$ at 1140 K.

DFT calculations of the aforementioned LiF, NaF, KF, and $CrF_3$ are used as the reference states to describe thermodynamic properties of the nine ternary compounds of $Li_3CrF_6$, $Na_3CrF_6$, $Na_5Cr_3F_{14}$, $NaCrF_4$, $K_3CrF_6$, $K_2CrF_5$, $KCrF_4$, $K_2Cr_5F_{17}$, and $Cr_2F_5$. Table 5 shows the present DFT values of formation enthalpy ($\Delta_f H_m$) of these ternary compounds using the LDA+U approach, together with the reactions to form these compounds. Table 5 shows that the $\Delta_f H_m$ values are negative for all ternary compounds with reference to their corresponding binary compounds. Yin et al. [16,36] conducted DFT calculations for $Li_3CrF_6$ and $Na_3CrF_6$. Predicted $\Delta_f H_m$ values from the Materials Project [78] and the Open Quantum Materials Database (OQMD) [79] are also listed in Table 5 and displayed in Figure 4. The present DFT calculations by LDA+U predict higher $\Delta_f H_m$ values of $Li_3CrF_6$ and $Na_3CrF_6$ than those by Yin et al. [16,36] using GGA. The present DFT calculations align better with results from the Materials Project [78] and OQMD [79] than calculations from Yin et al. [16,36]. Figure 4 displays the convex hulls for the LiF-$CrF_3$, NaF-$CrF_3$, and KF-$CrF_3$ systems based on the $\Delta_f H_m$ values listed in Table 5. These convex hulls serve as indicators regarding the stability of ternary compounds in these systems. $Li_3CrF_6$ is on the convex hull, suggesting that it is stable in the LiF-$CrF_3$ system. In the NaF-$CrF_3$ system, $Na_3CrF_6$ is located on the hull, indicating its stability at 0 K. $Na_5Cr_3F_{14}$ shows an elevation of 1.09 kJ/mol-atom above the hull, and $NaCrF_4$ shows 0.42 kJ/mol-atom above hull. In the KF-$CrF_3$ system, $K_2CrF_5$ is on the convex hull at 0 K. $K_2Cr_5F_{17}$ is the farthest away from the convex hull (1.31 kJ/mol-atom above it), while $K_3CrF_6$ and $KCrF_4$ are 0.62 kJ/mol-atom and 0.61 kJ/mol-atom above the hull, respectively. These calculations at 0 K suggests that $Na_5Cr_3F_{14}$, $NaCrF_4$, $K_2Cr_5F_{17}$, $K_3CrF_6$, and $KCrF_4$ are not stable at 0 K, while De Kozak [27] reported the existence of above ternary compounds at high temperature. It suggests that phonon-based QHA is necessary to investigate the thermodynamic properties of ternary compounds at high temperatures.

Figure 5 shows the predicted heat capacities $C_p$ of ternary compounds in the AF-$CrF_3$ (A= Li, Na, and K) systems from the phonon-based QHA, in comparison with the results estimated by the Neumann-Kopp rule [37], which were used in CALPHAD modeling by Dumaire et al. [18]. It shows that the Neumann-Kopp rule gives a good match with results by phonon-based QHA in the LiF-$CrF_3$ system. However, in the NaF-$CrF_3$ and KF-$CrF_3$ systems, the Neumann-Kopp rule estimates higher values of $C_p$ with respect to the values predicted by phonon-based QHA. The differences between the LiF-$CrF_3$ system and the NaF/KF-$CrF_3$ may be attributed to variations in melting temperatures between ternary and corresponding binary compounds. For example, in the LiF-$CrF_3$ system, the melting temperature of $Li_3CrF_6$ is reported at 1129 K [27], which is close to that of LiF at 1121 K [56] and below that of $CrF_3$ at 1698 K [56]. Considering the temperature below the melting point of $Li_3CrF_6$ ($T < 1129$ K), there are reliable resources of $C_p$ data from two endmembers LiF and $CrF_3$, thus the Neumann-Kopp rule $C_p$ estimation of $Li_3CrF_6$ is acceptable. However, in the NaF-$CrF_3$ system, $Na_3CrF_6$ melts at 1413 K [27], while NaF melts at 1269 K [56], indicating that there is an approximately 150 K temperature range without reliable Cp data for NaF. In contrast, the phonon-based calculations are direct predictions of ternary compounds and it provides more accurate descriptions of thermodynamic properties for compounds than those by the Neumann-Kopp rule used by Dumaire et al. [18], especially at high temperatures. Therefore, the phonon-based QHA results were used in the present CALPHAD modeling to improve the accuracy in describing ternary compounds. Note that the $C_p$ values for compounds in the (LiF, NaF, KF, and $CrF_2$)-$CrF_3$ systems can be predicted using the Supplementary XML file.

## 4.2 Thermodynamic modeling

Figure 6 shows the phase diagrams of the (LiF, NaF, KF, and $CrF_2$)-$CrF_3$ systems calculated from present models and compared with experimental data by De Kozak [27,28] and Sturm [30]. The present CALPHAD modeling shows a good agreement regarding phase boundaries with experimental data. As an example, Table 4 summarizes the presently modeled parameters for liquid, while the complete thermodynamic database can be found in the Supplementary XML file. Details of the invariant reactions and the congruent melting temperature calculated from the present modeling work compared to experiments [27,28,30] and ML predictions using Hong et al.'s model [64] are listed in Table 6 with discussion below.

In the LiF-$CrF_3$ system, the present prediction of the eutectic reaction Liquid ↔ LiF + $Li_3CrF_6$ at $x(CrF_3) = 0.159$ and T = 1004 K matches well with experimental values of $x(CrF_3) = 0.150$ and T = 1003 K [27]. The largest difference is observed for Liquid ↔ $CrF_3$ + $Li_3CrF_6$, with eutectic temperature of 1051 K from the present modeling, which is 8 K lower than the 1059 K reported by De Kozak [27]. The eutectic composition for this reaction at $x(CrF_3) = 0.345$ is notably improved from the 0.363 by Dumaire et al. [18] in comparison with the measured value 0.350 by De Kozak [27]. The present prediction of congruent melting temperature (1134 K) is 5 K higher than experiment (1129 K by De Kozak) [27], improved from the prediction (1111 K) by Dumaire et al. [18].

In the NaF-$CrF_3$ system, the present modeling gives good predictions of eutectic and peritectic temperatures compared with experiments [27] with the largest difference of 3 K observed for Liquid + $CrF_3$ ↔ $NaCrF_4$ (1231 K from the present work and 1234 K by De Kozak [27]). The eutectic composition for the reaction Liquid ↔ NaF + $Na_3CrF_6$ is $x(CrF_3) = 0.093$, which is around 0.03 away from the experimental value $x(CrF_3) = 0.123$ [27]. The eutectic temperature for this reaction by the present modeling is 1164 K, whereas it is 1175 K modelled by Dumaire et al. [18]. In comparison to experimental value of 1166 K by De Kozak [27], the difference reduces from 9 K to 2 K. The present congruent melting temperature of $Na_3CrF_6$ is predicted at 1403 K, improving from the predicted 1385 K by Dumaire et al. [18] but slightly lower than experimental value of 1413 K [27]. For the two eutectic reactions without experimental data, i.e., Liquid ↔ $Na_5Cr_3F_{14}$ + $Na_3CrF_6$ and Liquid ↔ $Na_5Cr_3F_{14}$ + $NaCrF_4$, the present modeling provides similar predictions ($x(CrF_3) = 0.373$, T = 1139 K; $x(CrF_3) = 0.378$, T = 1144 K respectively) compared to those modeled by Dumaire et al. [18] ($x(CrF_3) = 0.371$, T = 1145 K; $x(CrF_3) = 0.383$, T = 1144 K, respectively).

In the KF-$CrF_3$ system, the present modeling work predicts the congruent melting of $K_3CrF_6$ at 1555 K, which is in good agreement with the experimental value of 1553 K [27]. For the eutectic reaction, Liquid ↔ $KCrF_4$ + $K_2CrF_5$, the present modeling predicts $x(CrF_3) = 0.408$ and $T = 1103$ K, whereas the values are $x(CrF_3) = 0.45$ and $T = 1112$ K modeled by De Kozak [27]. The present modeling improves the prediction of eutectic temperature of Liquid ↔ KF + $K_3CrF_6$ to 1110 K from the predicted 1108 K by Dumaire et al. [18], compared experimental value of 1115 K by De Kozak [27]. For the three peritectic reactions in the KF-$CrF_3$ system, the present modeling work predicts 1141 K for Liquid + $K_3CrF_6$ ↔ $K_2CrF_5$, 1194 K for Liquid + $K_2Cr_5F_{17}$ ↔ $KCrF_4$, and 1394 K for Liquid + $CrF_3$ ↔ $K_2Cr_5F_{17}$. These temperatures are comparable to 1133 K, 1200 K, and 1390 K, respectively, as measured by De Kozak [27], showing an MAE of 6 K.

In the $CrF_2$-$CrF_3$ system, the present modeling work improves the predicted invariant compositions. For the eutectic reaction Liquid↔$CrF_2$+$Cr_2F_5$, the present work predicts the eutectic point at $x(CrF_3) = 0.124$. This value is higher than $x(CrF_3) = 0.115$ by Dumaire et al. [18] but aligns more closely with the experimental value $x(CrF_3) = 0.14$ by Sturm [30]. The eutectic temperature for this reaction is 1101 K predicted by the present modeling, which remains close to experimental 1103 K by Sturm [30]. For the peritectic reaction Liquid + $CrF_3$ ↔ $Cr_2F_5$, the present work predicts $x(CrF_3) = 0.299$ compared to measured $x(CrF_3) = 0.29$ by Sturm [30]. The present invariant temperature is predicted at 1276 K, demonstrating a slightly higher value by 4 K than the measured 1272 K by Sturm [30]. In the present work, the single-phase region of the $Cr_2F_5$ solid solution phase ranges from $x(CrF_3) = 0.38$ to $x(CrF_3) = 0.46$. This range aligns better with the suggested values of x($CrF_3$) from 0.40 to 0.45 by Sturm [30], than x($CrF_3$) from 0.382 to 0.45 by Dumaire et al.[18]. Overall, by incorporating thermodynamic data of compounds by the present DFT calculations, the present modeling work yields improved predictions of phase diagrams.

In addition, the present CALPHAD modeling work implements the mixing enthalpy of liquid at 1700 K, which was obtained by the present AIMD simulations as described in Sec.3.2. Figure 7 shows the mixing enthalpy of liquid at 1700 K by AIMD at different compositions compared to the present modeling results and results by Dumaire et al. [18]. It clearly shows that the mixing enthalpy of liquid (dot-dashed lines) by Dumaire et al.'s modeling [18] is much less negative compared to the present AIMD calculations. For example, in the LiF-$CrF_3$ system at x($CrF_3$) = 0.2, AIMD predicts the mixing enthalpy of -20.60 kJ/mol-atom at 1700 K, while Dumaire et al's modeling [18] shows -12.65 kJ/mol-atom, representing a 39% higher value. Using the present AIMD data of liquid for modeling, the present modeling work (solid lines) shows a great improvement. The present modeling work improves the prediction to -20.13 kJ/mol-atom at x($CrF_3$) = 0.2 for the LiF-$CrF_3$ system. The differences between the AIMD results and Dumaire et al.'s work [18] are more pronounced in the NaF-$CrF_3$ and KF-$CrF_3$ systems. In the NaF-$CrF_3$ system, at x($CrF_3$) = 0.333 (a composition region near the lowest mixing enthalpy), Dumaire et al.'s modeling [18] shows the mixing enthalpy of -25.45 kJ/mol-atom, while our modeling predicts -45.68 kJ/mol-atom. At the same condition, AIMD gives the mixing enthalpy of -47.37 kJ/mol-atom, which is about 22 kJ/mol-atom lower than Dumaire et al.'s results [18]. In the KF-$CrF_3$ system at x($CrF_3$) = 0.333, Dumaire et al. [18] predicts -30.58 kJ/mol-atom, significantly higher than -50.20 kJ/mol-atom by AIMD. The present modeling predicts -47.41 kJ/mol-atom, reducing the difference from the above mentioned 39% to the present 6%. It highlights that the present AIMD simulations enhances the reliability of the present modeling in describing liquid compared to the previous work [18]. The mixing enthalpy (dotted lines) modeling by Yin et al. [16] in the NaF-$CrF_3$ and KF-$CrF_3$ systems agree well with the results by the present modeling. Near the low mixing enthalpy region x($CrF_3$) = 0.333 in the NaF-$CrF_3$ system, Yin et al. [16] suggests the mixing enthalpy of -40.54 kJ/atom, which is 10% higher than -45.68 kJ/atom from the present work. Note that Yin et al. [16] used an empirical model proposed by Robelin and Chartrand [80] to estimate the mixing enthalpy in liquid. Overall, the present modeling work has improved the predictions of liquid than the modeling works by Dumaire et al. [18] and Yin et al. [16].

In Yin et al.'s modeling work [16], the associate model was used to describe the liquid phase. This model was applied to describe the short-range ordering (SRO) by assuming 'associates', such as $K_3CrF_6$ and $Na_3CrF_6$ associates. This kind of assumption may cause issues when extrapolation into higher order systems [81]. In the present work, the MQMQA was employed to describe liquid and provides information of the first and the second nearest neighbors in complex liquid. As an example, Figure 8 shows the predicted fraction of each quadruplet in the liquid phase. The composition where the peak fraction of the ACr/FF (A=Li, Na, and K) quadruplets appears, indicates the SRO. In the LiF-$CrF_3$ system, the peak fraction of LiCr/FF with strong SRO is around x($CrF_3$) = 0.25, which is consistent with the lowest mixing enthalpy around the x($CrF_3$) = 0.25 as shown in Figure 7. In addition, Figure 8 presents the quadruplet fractions and the neighboring environments of ions in liquid, which are difficult to obtain from the associate model due to its focus on associate clusters.

## 5 Conclusions

The present work revisits thermodynamic properties of compounds and liquid in the (LiF, NaF, KF, $CrF_2$)-$CrF_3$ systems by utilizing CALPHAD modeling with inputs from the DFT-based first-principles, phonon, and AIMD calculations. Thermodynamic properties, including enthalpy, entropy, and heat capacity of the binary (endmember) compounds LiF, NaF, KF, $CrF_3$, and $CrF_2$ as a function of temperature, have been predicted by DFT-based phonon calculations, agreeing with available experimental data in the literature and validating the reliability of the present methodology. They enabled the remodeling of the (LiF, NaF, KF, $CrF_2$)-$CrF_3$ systems with more accurate inputs. The MQMQA is employed to describe the liquid phase, providing valuable insights into the complex nature of molten salts such as the short-range ordering and neighboring of cations. Phase equilibria from the present CALPHAD modeling match better with experimental data in comparison with previous modeling work in the literature. The present thermodynamic data, including equilibrium volumes, bulk moduli, enthalpies, entropies, and heat capacities of compounds in the (LiF, NaF, KF, $CrF_2$)-$CrF_3$ system can be used to facilitate the development of advanced molten salt reactors.

## Acknowledgments

The authors acknowledge financial supports by the U.S. Department of Energy, Office of Nuclear Energy's Nuclear Energy University Programs via Award Nos. DE-NE0008945 and DE-NE0009288. First-principles calculations were performed partially on the Roar supercomputer at the Pennsylvania State University's Institute for Computational and Data Sciences (ICDS), partially on the resources of the National Energy Research Scientific Computing Center (NERSC) supported by the U.S. Department of Energy, Office of Science User Facility operated under Contract No. DE-AC02-05CH11231, and partially on the resources of ACCESS (previously the Extreme Science and Engineering Discovery Environment, XSEDE) supported by National Science Foundation (NSF) with Grant No. ACI-1548562.

## 6 Figures and Figure Captions

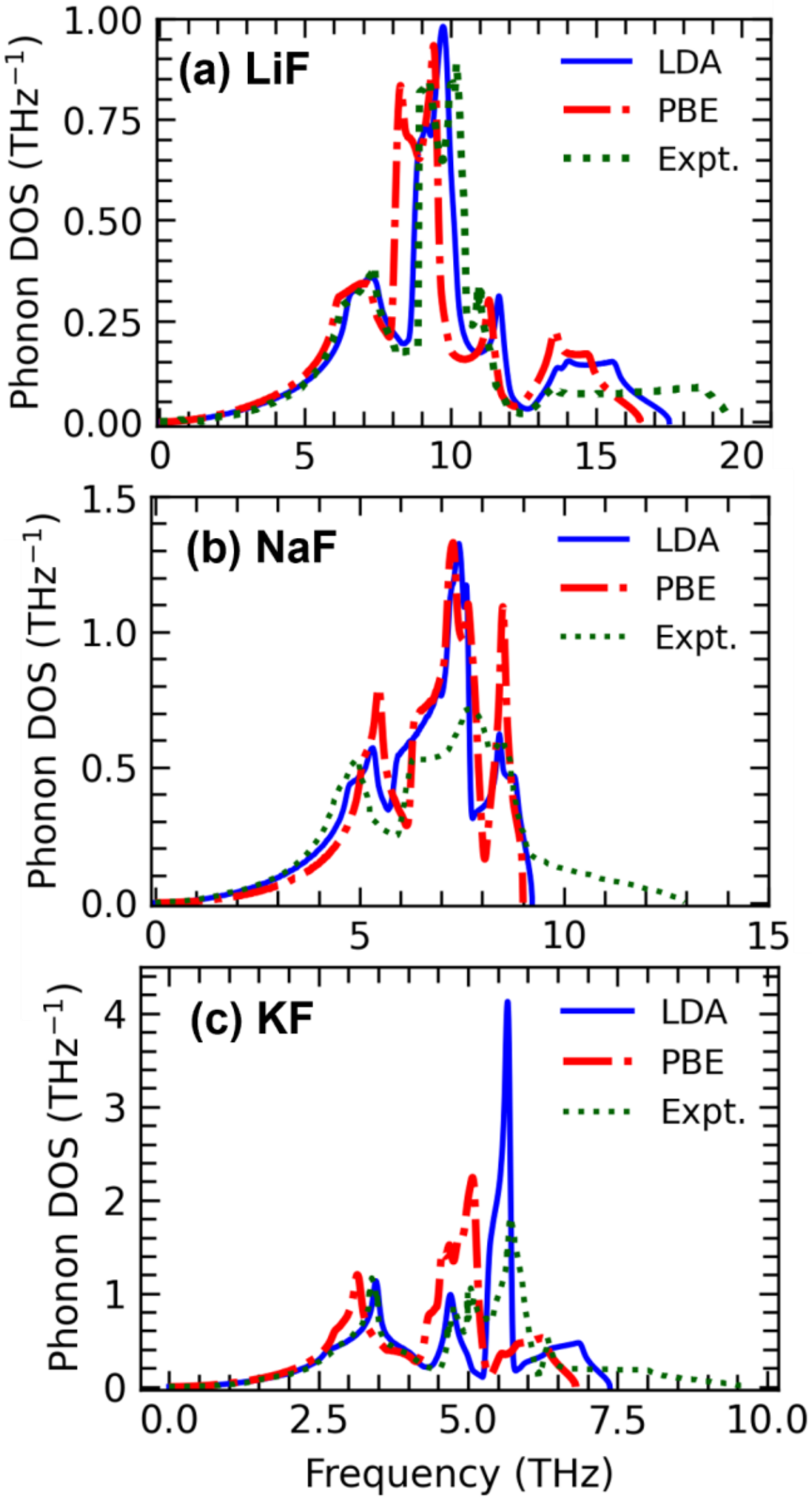

Figure 1. Predicted phonon density of states (DOS) of (a) LiF, (b) NaF, and (c) KF from DFT-based phonon calculations in comparison with phonon DOS from experiments [74–76]. Blue lines show results from the LDA approach, red dot-dashed lines show results from the GGA-PBE approach, and the green dotted lines are results from experiments [74–76].

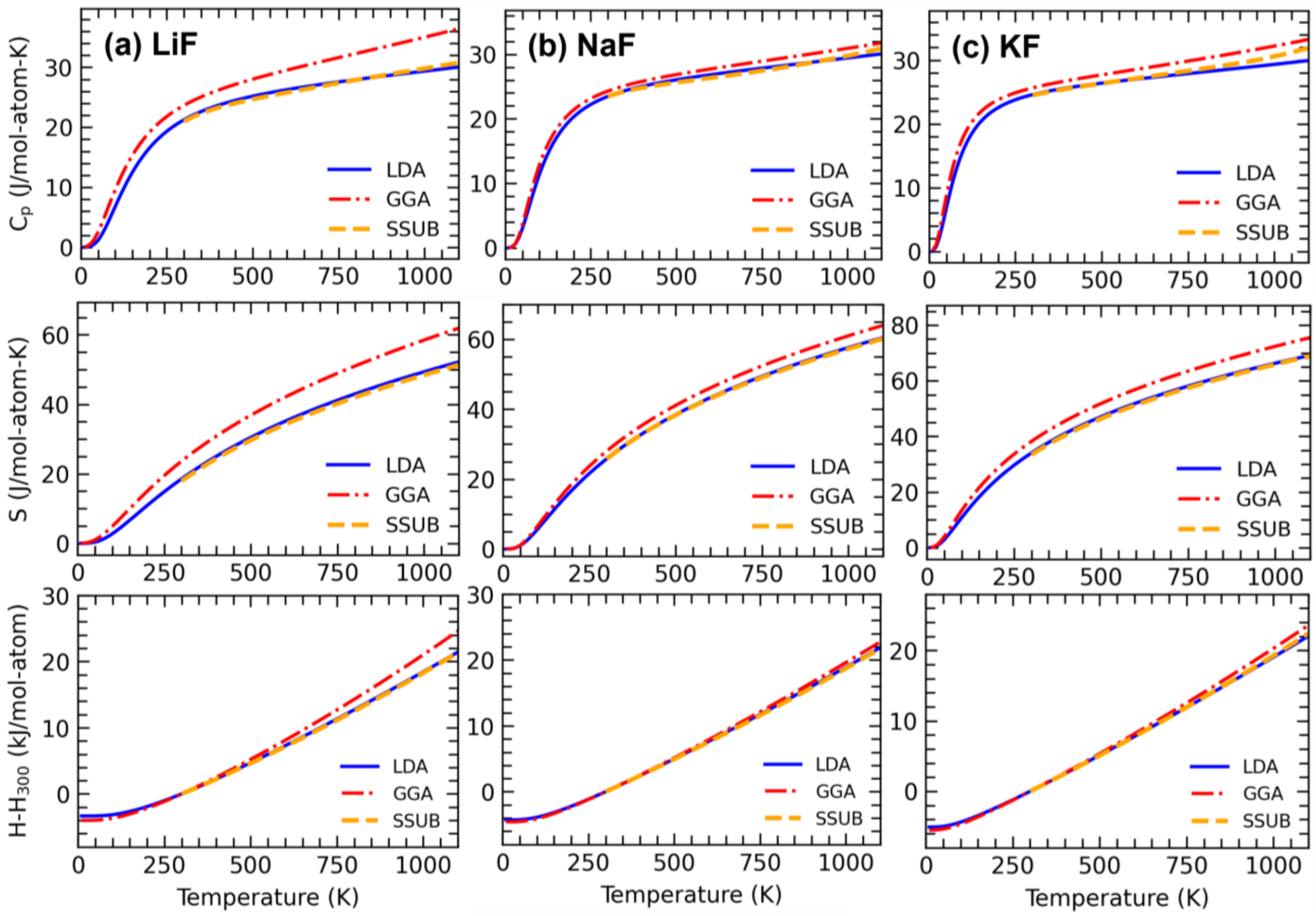

Figure 2. Predicted heat capacity $C_p$, entropy S, and enthalpy with reference at 300 K (H-$H_{300}$) of (a) LiF, (b) NaF, and (c) KF by DFT-based QHA via phonon calculations. Results by LDA is marked as solid blue lines, calculations by PBE are marked as dashed dot red lines, and the SSUB results [56] are in dashed yellow lines. The SSUB results are implemented in the present modeling work.

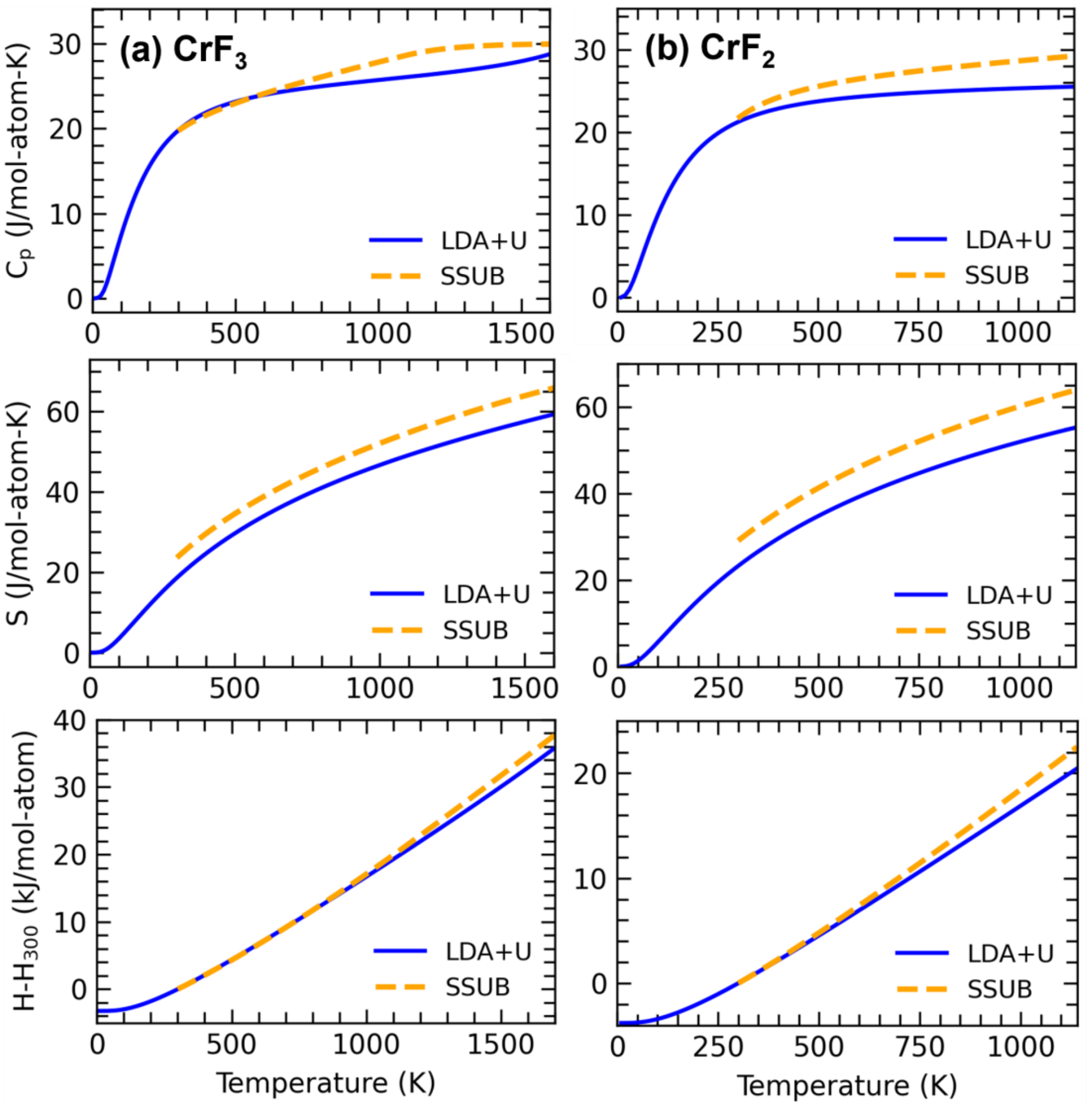

Figure 3. Predicted heat capacity $C_p$, entropy S, and enthalpy with reference at 300 K (H-$H_{300}$) of (a) $CrF_3$ and (b) $CrF_2$ by DFT-based QHA via phonon calculations marked in solid lines in comparison with SSUB [56] in dashed lines.

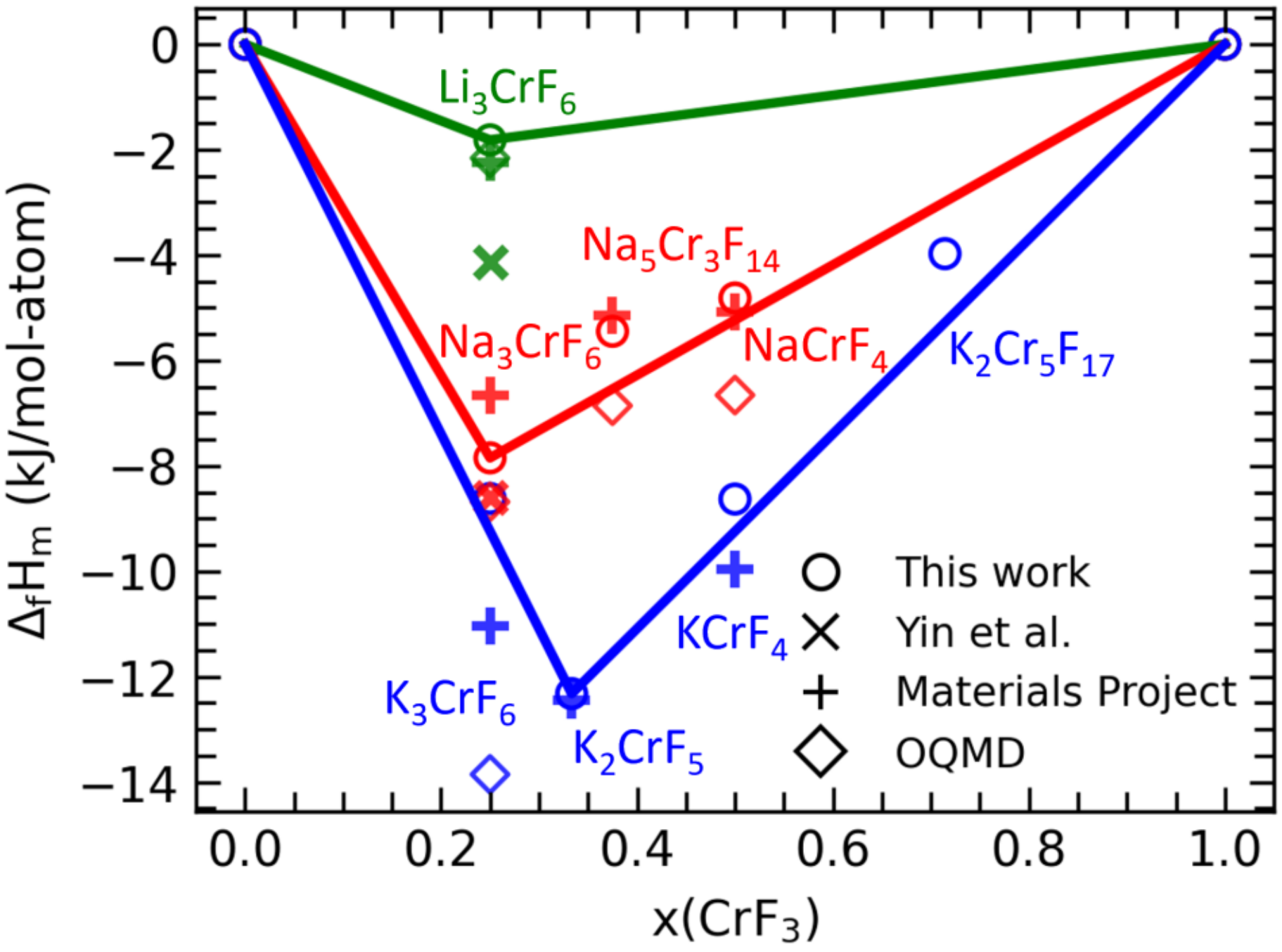

Figure 4. Convex hull of ternary compounds in LiF-$CrF_3$ (green), NaF-$CrF_3$ (red), and KF-$CrF_3$ (blue) at 0 K from DFT-based calculations from the present work. Circles (O) are formation enthalpy of compounds from the present work; cross markers (×) are formation enthalpy of compounds from Yin et al. [16,36]; plus markers (+) are results from The Materials Project [78]; and diamond markers (◊) are results from OQMD [79].

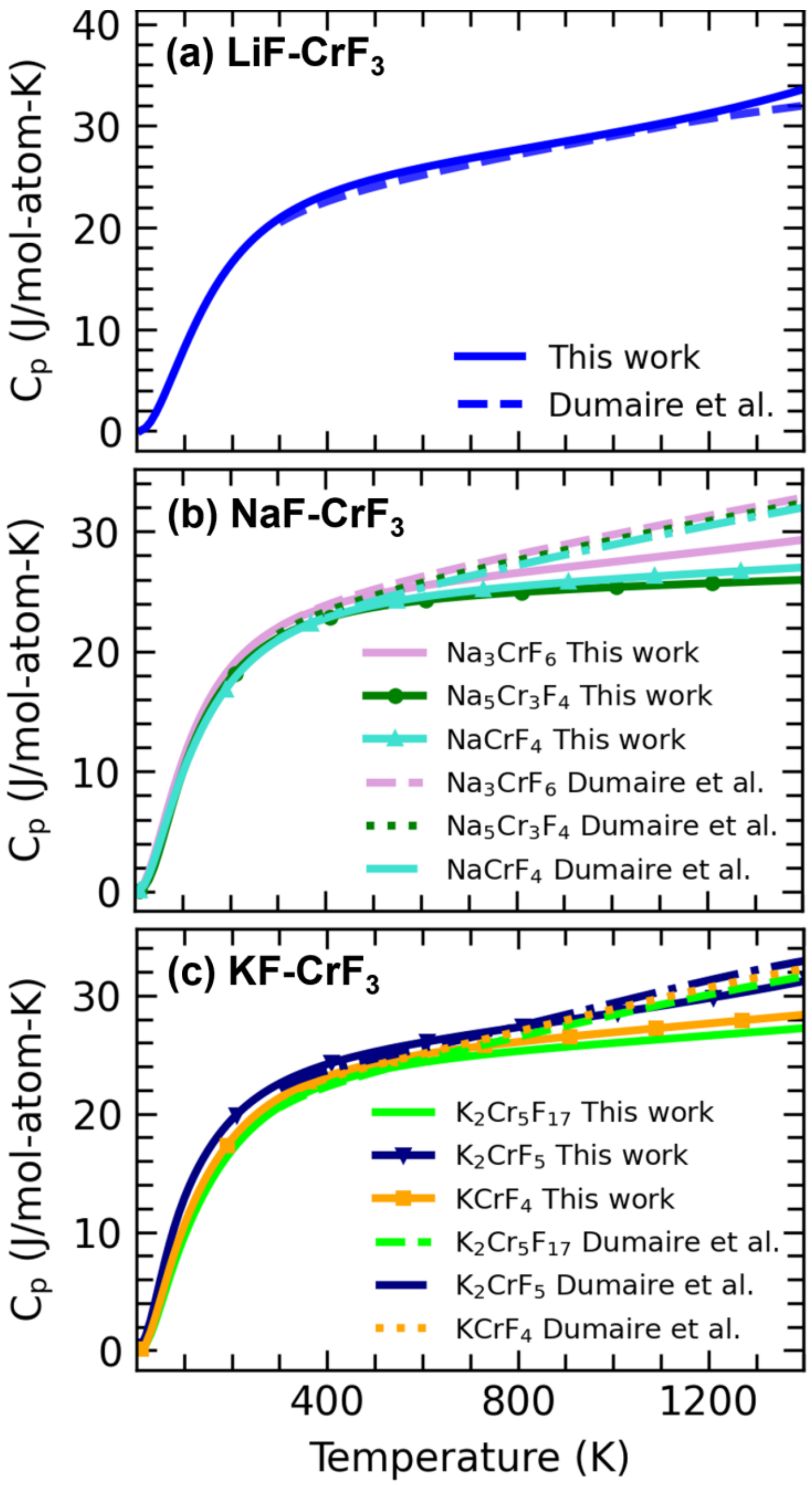

Figure 5. Predicted heat capacities of ternary compounds in the (a) LiF-$CrF_3$, (b) NaF-$CrF_3$, and (c) KF-$CrF_3$ systems by DFT-based QHA via phonon calculations (solid lines) compared with the Dumaire et al. [18]'s work (dashed lines). The QHA results are implemented in the present modeling work for ternary compounds.

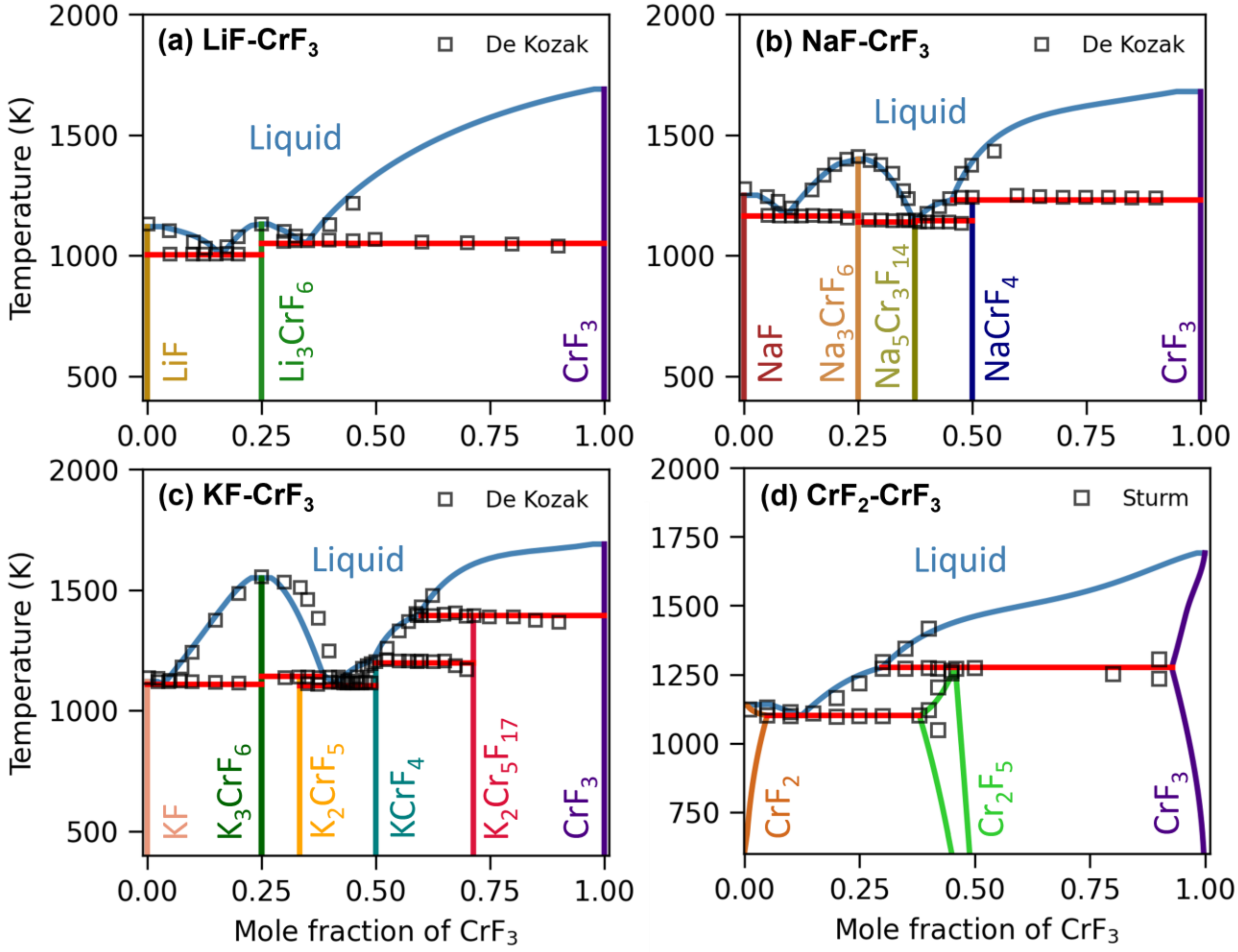

Figure 6. Predicted phase diagrams of the (a) $LiF$-$CrF_3$, (b) $NaF$-$CrF_3$, (c) $KF$-$CrF_3$, and (d) $CrF_2$-$CrF_3$ systems by the present CALPHAD modeling in comparison with experimental data [27,28,30].

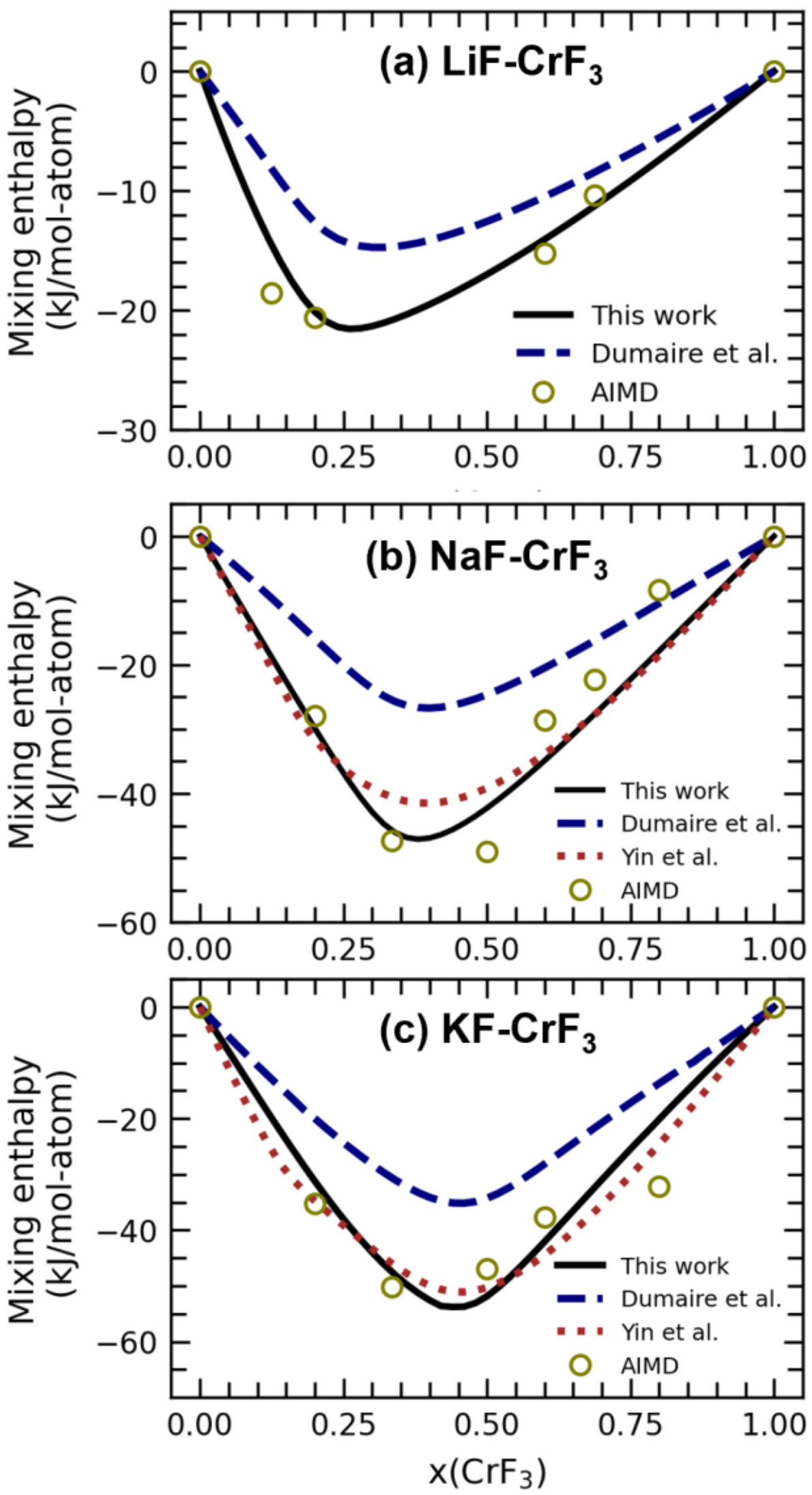

Figure 7. Predicted mixing enthalpy of liquid at 1700 K in (a) LiF-$CrF_3$, (b) NaF-$CrF_3$, and (c) KF-$CrF_3$ by the present CALPHAD modeling work (black solid lines), compared with the present AIMD results (circles) and modeling results by Dumaire et al. (blue dashed lines) [18] and Yin et al. (brown dotted lines) [16].

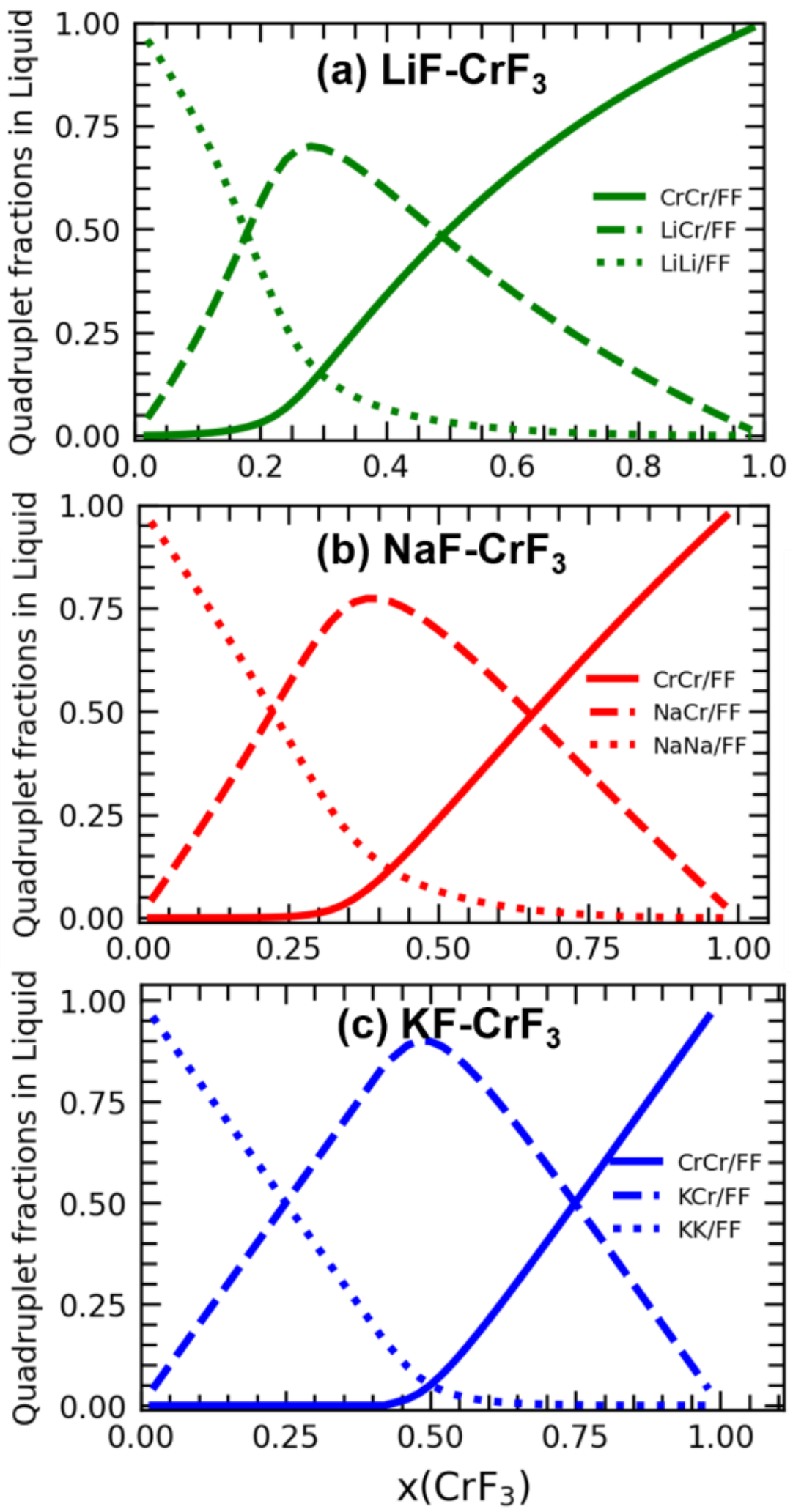

Figure 8. Precited quadruplet fractions in (a) LiF-$CrF_3$ (green lines), (b) NaF-$CrF_3$ (red lines), and (c) KF-$CrF_3$ (blue lines) liquid at 1700 K according to the present CALPHAD modeling.

## 7 Tables and Table Captions

Table 1. Details of DFT-based first-principles, phonon, and AIMD simulations for each compound or phase, including space group, total atoms in the supercells, k-point meshes for structure relaxations and the final static calculations (indicated by DFT), supercell sizes for phonon calculations, k-point meshes for phonon calculations, and k-point meshes for AIMD calculations.

| Phase | Space Group | Atoms in crystallographic cell | k-points for DFT | Atoms in supercell for phonon | k-points for phonon | k-points for AIMD |
|---|---|---|---|---|---|---|
| LiF | $Fm\bar{3}m$ | 8 | 10×10×10 | 32 | 10×10×10 | N/A |
| NaF | $Fm\bar{3}m$ | 8 | 10×10×10 | 32 | 10×10×10 | N/A |
| KF | $Fm\bar{3}m$ | 8 | 10×10×10 | 32 | 10×10×10 | N/A |
| $CrF_3$ | $R\bar{3}c$ | 24 | 9×9×3 | 24 | 9×9×3 | N/A |
| $CrF_2$ | $P2_1/c$ | 6 | 14×10×9 | 24 | 13×10×9 | N/A |
| $Li_3CrF_6$ | C2/c | 60 | 2×2×2 | 60 | 2×2×2 | N/A |
| $Na_3CrF_6$ | $P2_1/c$ | 20 | 9×8×5 | 40 | 9×8×5 | N/A |
| $Na_5Cr_3F_{14}$ | $P2_1/c$ | 44 | 6×6×3 | 44 | 6×6×3 | N/A |
| $NaCrF_4$ | $P2_1/c$ | 24 | 6×8×5 | 24 | 6×8× 6 | N/A |
| $K_3CrF_6$ | $Fm\bar{3}m$ | 40 | 5×5×5 | 40 | 5×5×5 | N/A |
| $K_2CrF_5$ | Pbcn | 128 | 3×1×1 | 128 | 3×1×1 | N/A |
| $KCrF_4$ | Pnma | 144 | 3×1×1 | 144 | 3×1×1 | N/A |
| $K_2Cr_5F_{17}$ | Cmcm | 96 | 2×2×2 | 96 | 2×2×2 | N/A |
| $Cr_2F_5$ | C2/c | 28 | 6×6×6 | N/A | N/A | N/A |
| Liquid | N/A | 108/128 | N/A | N/A | N/A | 1×1×1 |

Table 2. Coordination number used in the present CALPHAD modeling work with MQMQA for the liquid phase.

| A | B | $Z^{A}_{AB:FF}$ | $Z^{B}_{AB:FF}$ | $Z^{F}_{AB:FF}$ |
|---|---|---|---|---|
| $Li^{+}$ | $Li^{+}$ | 6 | 6 | 6 |
| $Na^{+}$ | $Na^{+}$ | 6 | 6 | 6 |
| $K^{+}$ | $K^{+}$ | 6 | 6 | 6 |
| $Cr^{3+}$ | $Cr^{3+}$ | 6 | 6 | 2 |
| $Li^{+}$ | $Cr^{3+}$ | 2 | 6 | 2 |
| $Na^{+}$ | $Cr^{3+}$ | 4 | 6 | 2.7 |
| $K^{+}$ | $Cr^{3+}$ | 6 | 6 | 3 |
| $Cr^{2+}$ | $Cr^{3+}$ | 6 | 6 | 2.4 |
| $Cr^{2+}$ | $Cr^{2+}$ | 6 | 6 | 3 |

Table 3. Predicted equilibrium properties of volume $V_0$, bulk modulus $B_0$, and first derivative of bulk modulus with respect to pressure $B'$ for each compound based on the present EOS fitting at 0 K (Eq. 2) in the FLiNaK-$CrF_3$-$CrF_2$ system. Experimental data (marked with *) [66–71] are also listed for comparison.

| Phases | Method | $V_0$ (Å$^3$/atom) | $B_0$ (GPa) | $B'$ |
|---|---|---|---|---|
| LiF | LDA | 7.488 | 86.5 | 4.25 |
| | GGA | 8.419 | 67.6 | 4.17 |
| | Yagi* [66] | | 66.5 | |
| | Haussühl* [67] | | 76.9 | |
| | Boehler et al.* [68] | | 65.4 | 4.98 |
| NaF | LDA | 11.457 | 61.4 | 4.74 |
| | GGA | 13.020 | 45.0 | 4.60 |
| | Yagi* [66] | | 45.9 | |
| | Haussühl* [67] | | 53.8 | |
| | Ramji Rao* [69] | | 52.3 | |
| | Bensch et al.* [70] | | 48.2 | 5.89 |
| KF | LDA | 17.363 | 43.5 | 5.06 |
| | GGA | 20.050 | 28.9 | 4.84 |
| | Yagi* [66] | | 37.0 | |
| | Haussühl* [67] | | 35.5 | |
| $CrF_3$ | LDA+U | 11.152 | 46.2 | 8.46 |
| | GGA+U | 12.902 | 29.3 | 7.29 |
| | Jørgensen et al.* [71] | | 29.2 | 10.3 |
| $CrF_2$ | LDA+U | 12.443 | 71.1 | 2.74 |
| $Li_3CrF_6$ | LDA+U | 9.543 | 56.1 | 5.85 |
| $Na_3CrF_6$ | LDA+U | 11.577 | 57.5 | 5.46 |
| $Na_5Cr_3F_{14}$ | LDA+U | 11.757 | 52.2 | 5.15 |
| $NaCrF_4$ | LDA+U | 11.933 | 53.0 | 4.35 |
| $K_3CrF_6$ | LDA+U | 15.705 | 51.5 | 5.67 |
| $K_2CrF_5$ | LDA+U | 13.858 | 45.9 | 5.65 |
| $KCrF_4$ | LDA+U | 13.920 | 38.1 | 4.88 |
| $K_2Cr_5F_{17}$ | LDA+U | 13.314 | 49.3 | 6.91 |
| $Cr_2F_5$ | LDA+U | 12.225 | 46.6 | 7.91 |

Table 4. Model parameters for excess Gibbs energy of the liquid phase in the (LiF, NaF, KF, $CrF_2$)-$CrF_3$ systems.

| System | Parameter values |
|---|---|
| LiF-$CrF_3$ | $\Delta g^{ex}_{\mathrm{LiCr:F_2}} = -31640.38 + 1.63T + (-13214.96\ +\ 14.99T)\chi_{\mathrm{LiCr/F_2}} + (-7028.37\ +\ 2.00T)\chi_{\mathrm{CrLi/F_2}} + (189.18 - 3.60T)\chi^2_{\mathrm{CrLi/F_2}}$ |
| NaF-$CrF_3$ | $\Delta g^{ex}_{\mathrm{NaCr:F_2}} = -48031.37\ +\ -3.06T + (-4558.94\ - 21.26T)\chi_{\frac{\mathrm{NaCr}}{\mathrm{F_2}}} + (3537.07 +\ 5.49T)\chi_{\frac{\mathrm{CrNa}}{\mathrm{F_2}}} + 2000.00\chi^2_{\mathrm{NaCr/F_2}}$ |
| KF-$CrF_3$ | $\Delta g^{ex}_{\mathrm{LiCr:F_2}} = -36789.86\ - 35.26T + (5405.72 +\ 6.46T)\chi_{\mathrm{KCr/F_2}} + (-39957.003\ - 40.69T)\chi_{\mathrm{CrK/F_2}} + (24251.79 + 12.21T)\chi^2_{\mathrm{CrK/F_2}}$ |
| $CrF_2$-$CrF_3$ | $\Delta g^{ex}_{\mathrm{Cr^{2+}Cr^{3+}:F_2}} = -9425.83\ +\ 9.15T + (2931.90\ - 15.29T)\chi_{\frac{\mathrm{Cr^{2+}Cr^{3+}}}{\mathrm{F_2}}} + (-6800.00 - 4.50T)\chi_{\mathrm{Cr^{3+}Cr^{2+}/F_2}} + (6450.00 - 5.35T)\chi^2_{\mathrm{Cr^{2+}Cr^{3+}/F_2}}$ |

Table 5. DFT-based results of formation enthalpy ($\Delta_f H_m$) at 0 K of ternary compounds in the AF-$CrF_3$ (A=Li, Na, and K) systems with the reference states as shown in the reactions. DFT results from Yin et al. [16,36], the Materials Project [78], and the Open Quantum Materials Database (OQMD) [79] are listed for comparison.

| Compound | Reaction | $\Delta_f H_m$ (J/mol-atom) | Source |
|---|---|---|---|
| $Li_3CrF_6$ | $Li_3CrF_6 = 3LiF + CrF_3$ | -1815 | This work |
| | | -4144 | Yin et al.[36] |
| | | -2238 | Materials Project [78] |
| | | -2161 | OQMD [79] |
| $Na_3CrF_6$ | $Na_3CrF_6 = 3NaF + CrF_3$ | -7849 | This work |
| | | -8592 | Yin et al.[16] |
| | | -6657 | Materials Project [78] |
| | | -8683 | OQMD [79] |
| $Na_5Cr_3F_{14}$ | $Na_5Cr_3F_{14} = 5NaF + 3CrF_3$ | -5447 | This work |
| | | -5148 | Materials Project [78] |
| | | -6859 | OQMD [79] |
| $NaCrF_4$ | $NaCrF_4 = NaF + CrF_3$ | -4816 | This work |
| | | -5071 | Materials Project [78] |
| | | -6657 | OQMD [79] |
| $K_3CrF_6$ | $K_3CrF_6 = 3KF + CrF_3$ | -8621 | This work |
| | | -11038 | Materials Project [78] |
| | | -13855 | OQMD [79] |
| $K_2CrF_5$ | $K_2CrF_5 = 2KF + CrF_3$ | -12322 | This work |
| | | -12446 | Materials Project [78] |
| $KCrF_4$ | $KCrF_4 = KF + CrF_3$ | -8631 | This work |
| | | -9970 | Materials Project [78] |
| $K_2Cr_5F_{17}$ | $K_2Cr_5F_{17} = 2KF + 5CrF_3$ | -3974 | This work |

Table 6. Predicted invariant equilibria in the AF-$CrF_3$ (A=Li, Na, K) systems by the present CALPHAD modeling, compared with experimental data from De Kozak [27] and Sturm [30] (marked with *), and other modeling works [16–18,36].

| Reaction | | x($CrF_3$) | Temperature (K) | Source |
|---|---|---|---|---|
| Eutectic | Liquid↔LiF+$Li_3CrF_6$ | 0.159 | 1004 | This work |
| | | 0.150 | 1003 | De Kozak [27] * |
| | | 0.136 | 1008 | Dumaire et al.[18] |
| | | 0.148 | 1003 | Yin et al.[36] |
| Congruent melting | Liquid↔$Li_3CrF_6$ | 0.25 | 1134 | This work |
| | | 0.25 | 1129 | De Kozak [27] * |
| | | 0.25 | 1114 | Hong et al.[64] |
| | | 0.25 | 1111 | Dumaire et al.[18] |
| | | 0.25 | 1125 | Yin et al.[36] |
| Eutectic | Liquid↔$CrF_3$+ $Li_3CrF_6$ | 0.345 | 1051 | This work |
| | | 0.350 | 1059 | De Kozak [27] * |
| | | 0.363 | 1062 | Dumaire et al.[18] |
| | | 0.354 | 1058 | Yin et al.[36] |
| Eutectic | Liquid↔NaF+$Na_3CrF_6$ | 0.093 | 1164 | This work |
| | | 0.123 | 1166 | De Kozak [27] * |
| | | 0.106 | 1175 | Dumaire et al.[18] |
| | | 0.114 | 1162 | Yin et al.[16] |
| Congruent melting | Liquid↔$Na_3CrF_6$ | 0.25 | 1403 | This work |
| | | 0.25 | 1413 | De Kozak [27] * |
| | | 0.25 | 1404 | Hong et al.[64] |
| | | 0.25 | 1385 | Dumaire et al.[18] |
| | | 0.25 | 1416 | Yin et al.[16] |
| Eutectic | Liquid↔$Na_5Cr_3F_{14}$+$Na_3CrF_6$ | 0.373 | 1139 | This work |
| | | 0.371 | 1145 | Dumaire et al.[18] |
| | | 0.367 | 1142 | Yin et al.[16] |
| Eutectic | Liquid↔$Na_5Cr_3F_{14}$+$NaCrF_4$ | 0.378 | 1144 | This work |
| | | 0.381 | 1144 | Dumaire et al.[18] |
| | | 0.383 | 1141 | Yin et al.[16] |
| Peritectic | Liquid+$CrF_3$↔$NaCrF_4$ | 0.5 | 1231 | This work |
| | | 0.5 | 1234 | De Kozak [27] * |
| | | 0.5 | 1232 | Dumaire et al.[18] |
| | | 0.5 | 1239 | Yin et al.[16] |
| Eutectic | Liquid↔KF+$K_3CrF_6$ | 0.037 | 1110 | This work |
| | | 0.048 | 1115 | De Kozak [27] * |
| | | 0.041 | 1108 | Dumaire et al.[18] |
| | | 0.045 | 1113 | Yin et al.[17] |
| Congruent melting | Liquid↔$K_3CrF_6$ | 0.25 | 1555 | This work |
| | | 0.25 | 1553 | De Kozak [27] * |
| | | 0.25 | 1520 | Hong et al.[64] |
| | | 0.25 | 1553 | Dumaire et al.[18] |
| | | 0.25 | 1548 | Yin et al.[17] |
| Peritectic | Liquid+$K_3CrF_6$↔$K_2CrF_5$ | 0.333 | 1141 | This work |
| | | 0.333 | 1133 | De Kozak [27] * |
| | | 0.333 | 1130 | Dumaire et al.[18] |
| | | 0.333 | 1135 | Yin et al.[17] |
| Eutectic | Liquid↔$KCrF_4$+$K_2CrF_5$ | 0.408 | 1103 | This work |
| | | 0.45 | 1112 | De Kozak [27] * |
| | | 0.432 | 1112 | Dumaire et al.[18] |
| | | 0.426 | 1107 | Yin et al.[17] |
| Peritectic | Liquid+$K_2Cr_5F_{17}$↔$KCrF_4$ | 0.5 | 1194 | This work |
| | | 0.5 | 1200 | De Kozak [27] * |
| | | 0.5 | 1191 | Dumaire et al.[18] |
| | | 0.5 | 1195 | Yin et al.[17] |
| Peritectic | Liquid+$CrF_3$↔$K_2Cr_5F_{17}$ | 0.714 | 1394 | This work |
| | | 0.714 | 1390 | De Kozak [27] * |
| | | 0.714 | 1390 | Dumaire et al.[18] |
| | | 0.714 | 1388 | Yin et al.[17] |
| Eutectic | Liquid↔$CrF_2$+$Cr_2F_5$ | 0.124 | 1101 | This work |
| | | 0.14 | 1103 | Sturm [30] * |
| | | 0.115 | 1104 | Dumaire et al.[18] |
| Peritectic | Liquid+$CrF_3$↔$Cr_2F_5$ | 0.299 | 1276 | This work |
| | | 0.29 | 1272 | Sturm [30] * |

| | | | |
|---|---|---|---|
| | 0.28 | 1271 | Dumaire et al.[18] |